\documentstyle[aps,prl,preprint,floats,epsfig]{revtex}

\newcommand{\pcite}{\protect\cite}

\begin{document}

\preprint{\tighten\vbox{\hbox{\hfil CLNS 99/1621}
                        \hbox{\hfil CLEO 99/5}
}}

\title{Limit on Tau Neutrino Mass from 
  $\tau^-\to\pi^-\pi^+\pi^-\pi^0\nu_\tau$}  

% Your author list ***DOES NOT*** go here!
% is goes below where you are instructed to insert it...
\author{CLEO Collaboration}
\date{\today}

\maketitle
\tighten

\begin{abstract} 
From a data sample of 29058 $\tau^\pm\to\pi^\pm\pi^+\pi^-\pi^0\nu_\tau$
decays observed in the CLEO detector we derive a 95\% confidence upper limit
on the tau neutrino mass of 28 MeV.
\end{abstract}
\pacs{14.60.Pq, 12.15.Ff, 23.40.Bw}
\newpage

{
\renewcommand{\thefootnote}{\fnsymbol{footnote}}

\begin{center}
M.~Athanas,$^{1}$ P.~Avery,$^{1}$ C.~D.~Jones,$^{1}$
M.~Lohner,$^{1}$ C.~Prescott,$^{1}$ A.~I.~Rubiera,$^{1}$
J.~Yelton,$^{1}$ J.~Zheng,$^{1}$
G.~Brandenburg,$^{2}$ R.~A.~Briere,$^{2}$
% ...
\end{center}
 
\small
\begin{center}
$^{1}${University of Florida, Gainesville, Florida 32611}\\
$^{2}${Harvard University, Cambridge, Massachusetts 02138}\\
% ...
$^{12}${University of Oklahoma, Norman, Oklahoma 73019}\\
% ...
$^{14}${University of Rochester, Rochester, New York 14627}\\
$^{15}${Stanford Linear Accelerator Center, Stanford University, Stanford,
California 94309}\\
% ...
$^{23}${University of California, Santa Barbara, California 93106}\\
% ...
\end{center}

\setcounter{footnote}{0}
}
\newpage

\section{Motivation}
In the Standard Model the tau lepton and the tau neutrino form the third
generation weak doublet of leptons.  Most observations are
consistent with zero mass for each of the three
types of neutrino, and with the conservation of lepton number for each of
$e,~\mu,~\tau$ separately.  These suppositions, however, should be tested,
especially in the light of recent indications
\cite{Rosc,RDCXXIV} of oscillations among the neutrino species.

There are model dependent limits on the possible values
of the tau neutrino mass.
The SuperKamiokande experiment \cite{RDCXXIV} measures the ratio of rates for
$\nu_\mu$ and $\nu_e$ from decay products of particles produced
in cosmic ray collisions in the atmosphere.  The deficit in the $\nu_\mu$
rate, along with its dependence upon neutrino zenith angle and energy, 
can be interpreted
as due to oscillation of the $\nu_\mu$ to $\nu_\tau$ with a maximal amplitude
($\sin 2\theta\sim 1$) and a frequency determined by
$5\times10^{-4}<\Delta m^2<6\times10^{-3}$ eV$^2$ at 90\% confidence.
This would imply, if the $\nu_\tau$ mass were
much larger than the $\nu_\mu$ mass, that
$$0.02<m_{\nu_\tau}<0.08~{\rm eV~~~(90\%~c.l.).}$$

Astrophysical observations and cosmological theory limit the energy
density of the universe, thus restricting the sum of {\it stable}
neutrino masses \cite{RCowsik}.  This leads to the limit
$$m_{\nu_\tau}< 24{\rm ~eV}.$$
This limit not only depends on $\nu_\tau$ being stable, but also varies with
the value of the Hubble constant and other inputs.

The possible effects of the tau neutrino on big-bang
nucleosynthesis lead to either a low mass $\nu_\tau$ or to a decaying
$\nu_\tau$ at higher mass
\cite{RKawa}:
$$m_{\nu_\tau}<0.37~{\rm MeV,~~or}~~m_{\nu_\tau}>18~{\rm MeV}.$$
The width of the excluded region increases with the lifetime of the $\nu_\tau$
and also depends on the assumed abundance of light elements.

The Standard Model relations among the $\tau$ and $\mu$ masses and lifetimes
and some $\tau$ branching fractions \cite{RDCXXVII} imply
$$m_{\nu_\tau}<48~{\rm MeV.}$$

The popular see-saw mechanism for generating neutrino masses \cite{RDCXV}
postulates the relation
$$m_{\nu_e}:m_{\nu_\mu}:m_{\nu_\tau}=m_e^2:m_\mu^2:m_\tau^2.$$
This would imply rather weak limits on the $\nu_\tau$ mass:
$$m_{\nu_\tau}<180~{\rm MeV,~~from~}m_{\nu_e}<15~{\rm eV},$$
$$m_{\nu_\tau}<48~{\rm MeV,~~from~}m_{\nu_\mu}<0.17~{\rm MeV}.$$

The model dependences in all of these limits on the $\nu_\tau$ mass argue for
a more direct measurement.  The observation of the decay of accelerator
produced taus along with measurements of the energy and momentum
of the detectable
daughter products (all but the $\nu_\tau$) can constrain the possible
values for $m_{\nu_\tau}$, especially in cases when the effective mass of the
detected particles is close to $m_\tau$.  Because the cross section for
$\tau^+\tau^-$ production is up to 15\% of the total $e^+e^-$ 
annihilation cross section, an electron-positron collider is a natural
choice for the source of taus.

One looks for a hadronic decay mode with only one neutrino in the
final state.  Most of the previous measurements 
(see Table \ref{Tlimits}) have been made with the higher multiplicity decay
modes in which the effective mass of the hadrons is more likely to be close
to the kinematic limit $m_\tau$
with the maximum sensitivity to $m_{\nu_\tau}$.  All such decay
modes are strongly phase-space suppressed, however, 
and the branching fractions are
very low.  An alternative strategy, which we use in the present measurement,
is to pick a decay mode with a lower hadron multiplicity but with a much
larger branching fraction.  Although the four-pion decay,
$\tau^-\to\pi^-\pi^+\pi^-\pi^0\nu_\tau$, relative to the five-pion
decay, produces a smaller proportion of events
in which the effective hadronic mass is close to $m_\tau$, the branching
fraction is 4.2\%, as compared to $\sim0.1\%$ for the higher multiplicity
modes.

%---------------------------------------------------------------------------
\section{Data Sample and Event Selection}
The experiment was performed using the Cornell Electron-positron Storage
Ring and the CLEO II detector, described elsewhere
\cite{Rdet}.  Charged particle tracks were reconstructed in three nested
cylindrical drift chambers in a solenoid field of 1.5 Tesla.
The mean-squared resolution in momentum transverse
to the beam was 
$(\delta p_T/p_T)^2=0.005^2+(0.0015{\rm~GeV}^{-1}\times p_T)^2$.
Photon and electron showers were detected over 98\% of $4\pi$ steradians
in an array of 7800 CsI scintillation counters with an energy resolution of 
$\delta E/E=0.0035/E^{0.75}+0.019-0.001E$ ($E$ in GeV) in the central region
of polar angle, $45^\circ<\theta<135^\circ$.
Ionization, time of flight, and shower energy aided in lepton identification.

The data used in the present analysis were from 4.75 fb$^{-1}$ of
accumulated luminosity, two thirds at
10.58 GeV and one third at 10.52 GeV $e^+e^-$ center of mass energies.
This corresponds to 4.3 million $\tau^+\tau^-$ pairs produced.  

We determined event selection criteria using Monte Carlo simulated signal
and background data samples.  
We selected the one-versus-three track topology
with zero net charge, that is, events containing a three-charged-track tau
decay candidate tagged by a single-prong decay in the opposite hemisphere.
Tracks were accepted in the polar angle range $|\cos\theta|<0.9$.
The three signal tracks each had to have $p_T>0.019E_{beam}$ and had to
fail electron identification criteria \cite{Rthesis}.
The tag track had to have momentum greater than $0.047E_{beam}$ and had to
be consistent with one of four possible decay modes:
$$\tau^+\to e^+\nu_e\overline{\nu_\tau},$$
$$\tau^+\to\mu^+\nu_\mu\overline{\nu_\tau},$$
$$\tau^+\to\pi^+\overline{\nu_\tau},$$
$$\tau^+\to\rho^+\overline{\nu_\tau},$$
or the charge conjugates.
This resulted in a data sample of $813$ thousand tagged three-prong tau decay
candidates.

On the three-prong side each event was required to have a $\pi^0$,
defined as two CsI calorimeter showers in the polar angle range
$|\cos\theta|<0.71$, not matched to charged tracks,
with lateral shower profiles consistent with
photons, and with effective mass in the range
$120<m_{\gamma\gamma}<150$ MeV.  The $\pi^0$ in the decay of the
$\rho^+$ in the $\tau^+\to\rho^+\overline{\nu_\tau}$ tag channel also had to
satisfy these requirements, along with $E_{\gamma}>50$ MeV for each photon
and $m(\pi^+\pi^0)<1.5$ GeV.
Non-photon calorimeter showers can make false $\pi^0$ candidates.
These typically originate from nuclear interactions of the
charged pions in the CsI crystals producing secondaries isolated from any
charged track.
If several $\pi^0$ candidates were found on the three-prong side of
an event, we kept only the one with the highest energy.  We rejected events
with an extra shower of more than 300 MeV or, if photon-like in lateral
shower shape, of more than 100 MeV.
The $\pi^0$ selection cuts reduced the data sample to 31,305 events.

In order to minimize background from the two-photon process, for example
$e^+e^-\to e^+e^-\gamma^*\gamma^*,~\gamma^*\gamma^*\to\pi^+\pi^+\pi^-\pi^-
\pi^0$ with the final $e^+$ and $e^-$ escaping detection at small 
angles to the
beam, we rejected events in which the net event momentum transverse to the
beam was less than 150 MeV.  The final data set included 29,058 events.

No particle identification information was used on the charged particles.
That is, decay modes in which a $K^\pm$ substitutes for a $\pi^\pm$
were considered part of the $\tau^-\to h^- h^+ h^-\pi^0\nu_\tau$
signal.  These modes contributed about 5\% of the signal
(see Table \ref{Tmodes}).
The four-pion final state includes $K_S^0\pi^-\pi^0\nu_\tau,~
K_S^0\to\pi^+\pi^-$, at 3\% of the total.

\begin{figure}
\epsfig{file=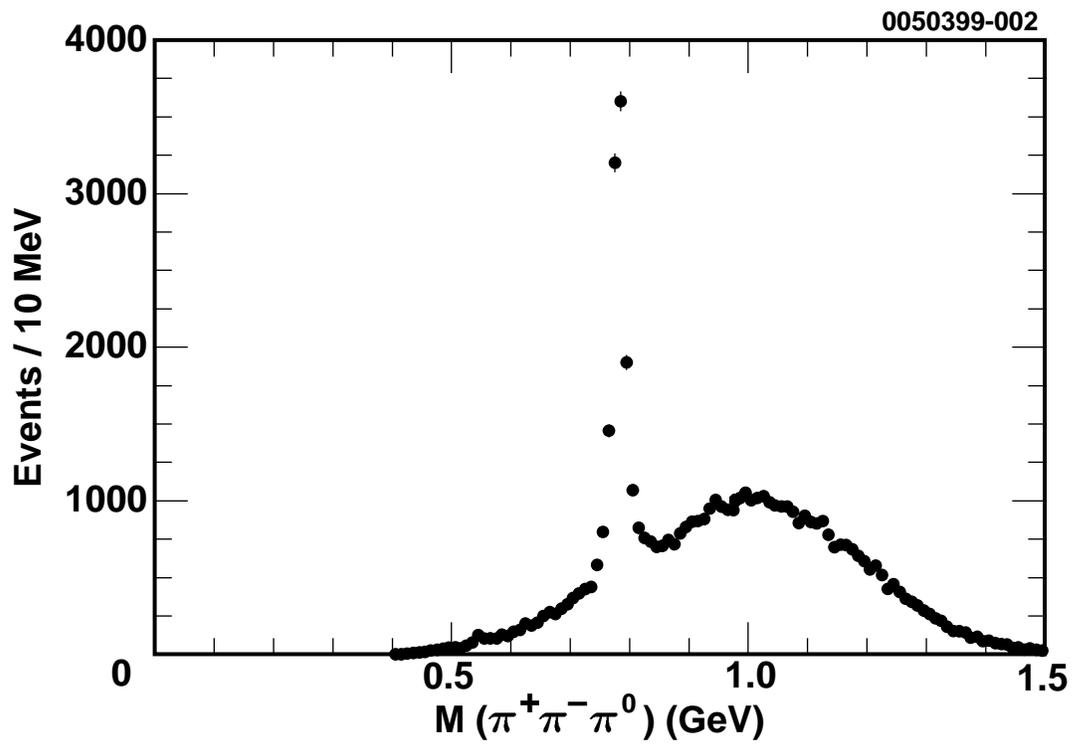,bbllx=90pt,bblly=420pt,bburx=510pt,
 bbury=700pt}
\caption{Invariant mass distribution for $\pi^+\pi^-\pi^0$ combinations
in data, with two entries per event.  No backgrounds have been subtracted.}
\label{Fomega}\end{figure}

\begin{figure}
\epsfig{file=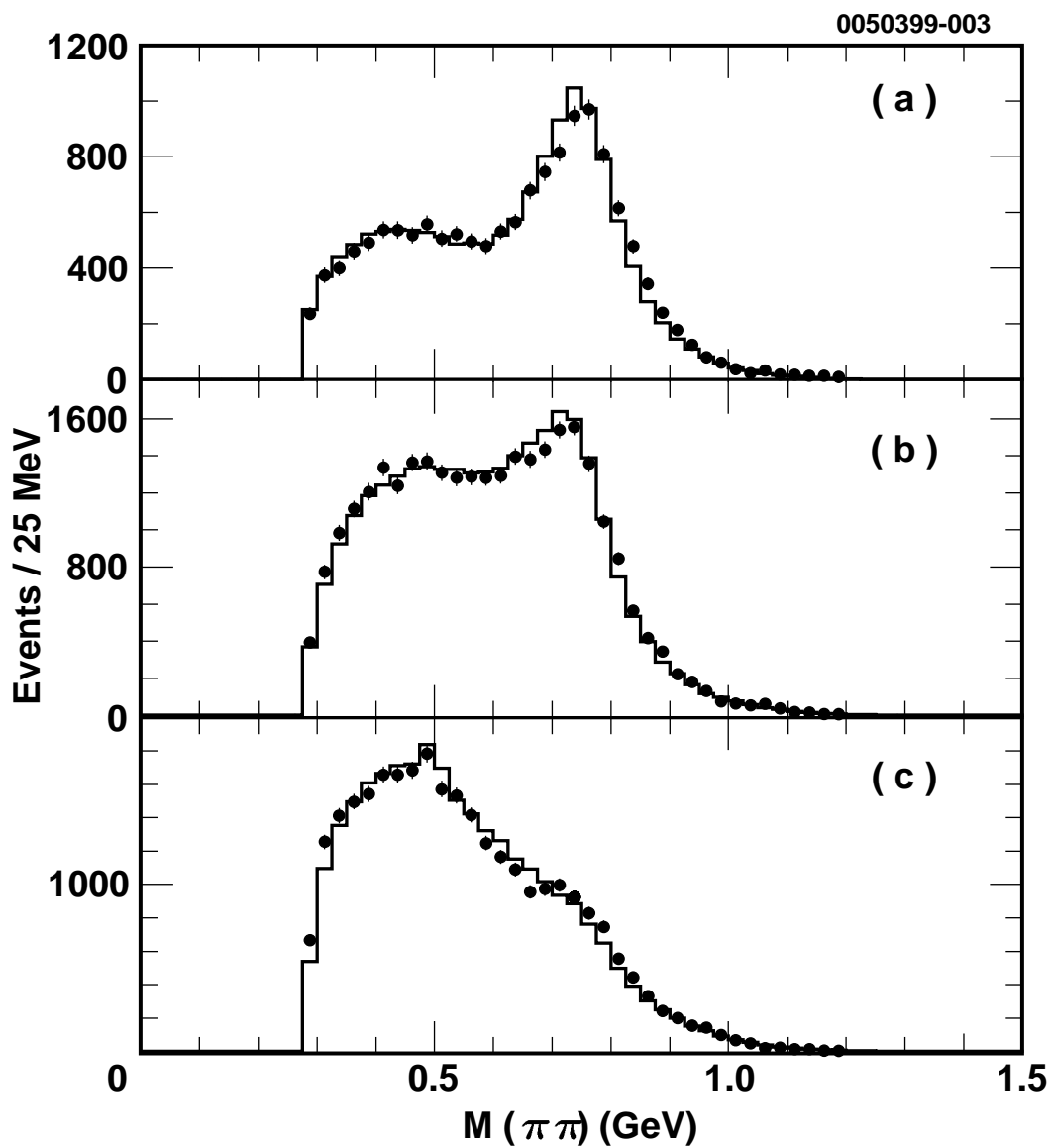,bbllx=90pt,bblly=190pt,bburx=510pt,
 bbury=660pt}
\caption{Invariant mass of $\pi\pi$ combinations in events outside the
$\omega$ peak: $\pi^+\pi^0$ (a, one entry per event), $\pi^-\pi^0$
(b, two entries per event), $\pi^+\pi^-$ (c, two entries per event).
The data distribution is represented by the points
with error bars and Monte Carlo by the histogram.}
\label{Ftwopi}\end{figure}

\begin{figure}
\epsfig{file=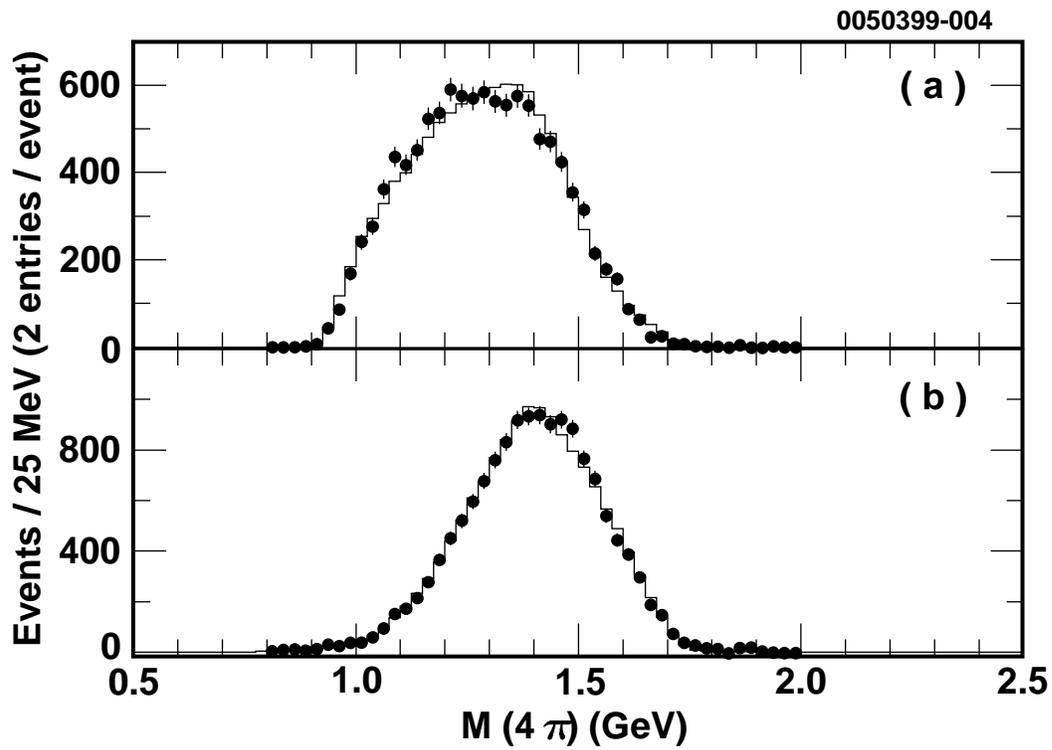,bbllx=90pt,bblly=390pt,bburx=510pt,
 bbury=690pt}
\caption{Four-pion invariant mass for events within the $\omega$ peak (a)
and outside the $\omega$ peak (b).  
The data distribution is represented by the points
with error bars and Monte Carlo by the histogram.
}

\label{Ffourpi}\end{figure}

The bulk of the signal is $\tau^-\to\pi^-\pi^+\pi^-\pi^0\nu_\tau$ (and charge
conjugate).  In about 53\% of the events there is a $\pi^+\pi^-\pi^0$
combination (Fig.\ \ref{Fomega}) with a mass consistent with the $\omega$
(there are two possibilities per event).  The two-pion mass spectra 
(Fig.\ \ref{Ftwopi}) for
events with no $\omega$ show $\rho$ peaks in  $\pi^+\pi^0$ ($21\%$ per event),
$\pi^-\pi^0$ (17\% per event), and $\pi^+\pi^-$ ($<2\%$ per event).
The overall four-pion mass spectrum (Fig.\ \ref{Ffourpi}) has a broad maximum
around 1.2 to 1.4 GeV for the $\omega\pi^-$ events, and a peak at 1.4 GeV
for the rest of the events.  There is no obvious resonance structure in
$m_{4\pi}$,
although the two four-pion spectra fit well each to a sum of $\rho(770),~
\rho(1450)$, and $\rho(1700)$ with adjustable relative amplitudes and phases.

%----------------------------------------------------------------------------

\section{Analysis}
\begin{figure}
\epsfig{file=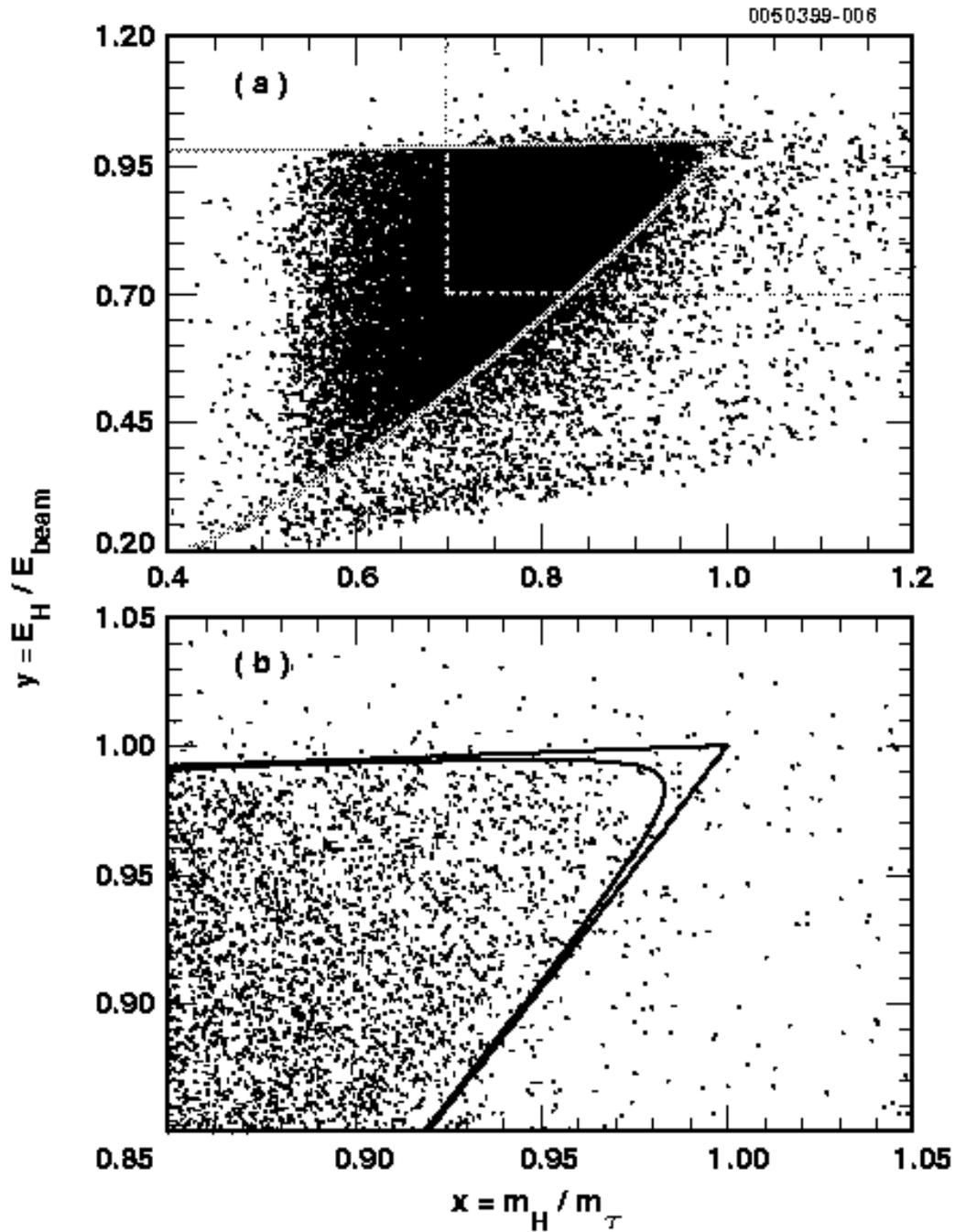,bbllx=90pt,bblly=140pt,bburx=510pt,
 bbury=710pt}
\caption{Distribution in scaled hadronic energy versus scaled hadronic mass
for events in the final data sample, shown for the full range (a) and for
a restricted range near the kinematic end point (b).
The curves show the boundaries of the allowed kinematic region for
$\tau\to\pi^-\pi^+\pi^-\pi^0\nu_\tau$ assuming $m_{\nu_\tau}=0$ in (a) and (b)
and $m_{\nu_\tau}=30$ MeV in (b).  The dashed lines in (a) show the boundary
of the fit region.}
\label{Fxy}\end{figure}

Conservation of energy and momentum imply that
\begin{eqnarray*}
m_{\nu_\tau}^2&=&(E_{beam}-E_H)^2-(\vec{p}_\tau-\vec{p}_H)^2 \\
       &=&m_\tau^2+m_H^2-2E_{beam}E_H+2\sqrt{E_{beam}^2
             -m_\tau^2}\sqrt{E_H^2-m_H^2}\cos\theta_{H\tau}.
\end{eqnarray*}
The beam energy $E_{beam}$ and the energy $E_H$ and effective mass $m_H$ 
of the hadronic (four-pion) system are measured in each event, and 
$m_\tau=1777.05^{+0.29}_{-0.26}$ MeV is known \cite{Rpdg,Rtaum}.  
Thus if we fix $m_{\nu_\tau}$,
for each allowed value of the scaled hadronic mass $x=m_H/m_\tau$ there
is a range of kinematically
allowed values of the scaled hadronic energy $y=E_H/E_{beam}$,
where the limits are obtained by taking the last term to be at its
$\cos\theta_{H\tau}=\pm1$ limits.  Figure~\ref{Fxy} shows the distribution
of the data in $x,y$ and the boundary curves for two values of
$m_{\nu_\tau}$. 
Even though there is background outside the kinematically allowed region,
it is clear that the two-dimensional distribution of the data is sensitive
to the value of $m_{\nu_\tau}$.  
More precisely, the likelihood of the observed
$x,y$ event distribution, including background, plotted for various assumed
$m_{\nu_\tau}$ values, can give information on which $m_{\nu_\tau}$ values
are consistent with experiment.

We define the likelihood for an individual event observed at $x_i,~y_i$
as the probability density $P(x_i,y_i|m_{\nu_\tau})$ 
of observing such an event
assuming $m_{\nu_\tau}$ to be the neutrino mass.   $P(x_i,y_i|m_{\nu_\tau})$
contains terms for signal and backgrounds.  The likelihood for the entire
data sample is then the product of the event likelihoods:
$${\cal L}(m_{\nu_\tau})=\prod_{i=1}^{N}P(x_i,y_i|m_{\nu_\tau}).$$
%.............................................

{\it Signal Likelihood.}
We first discuss the signal contribution to the single-event probability
density $P(x_i,y_i|m_{\nu_\tau})$.  It can be expressed as the
product of the spectral function (decay probability density) ${\cal F}$
and detection efficiency $\epsilon$ at the true $x,y$,
convolved with the experimental
resolution function ${\cal R}_i$ derived from data for that event:
$$P_{sig}(x_i,y_i|m_{\nu_\tau})=\int{\cal F}(x,y|0)w(x,y|m_{\nu_\tau})
 \epsilon(x,y){\cal R}_i(x_i-x,y_i-y)dxdy.$$
For convenience we have expressed ${\cal F}(x,y|m_{\nu_\tau})$ 
in terms of the
spectral function for $m_{\nu_\tau}=0$ and a weight function 
that takes account
of the dependence of ${\cal F}$ on $m_{\nu_\tau}$.
The weight function $w$ is determined from the known effect of a non-zero
neutrino mass on the phase space and the kinematic boundary.
It is zero outside the allowed region.

The spectral function ${\cal F}(x,y|0)$ is obtained by adjusting
a physics motivated 14-parameter function \cite{Rthesis}
to match distributions in the simulated data and the real data
over the range $x<0.925$, where we have verified by Monte Carlo that
the choice of ${\cal F}$ does not bias the determination of $m_{\nu_\tau}$.
The function so determined is then used in the entire $x<1$ range.  
It includes
adjustable amplitudes and phases for $\omega\pi$ and for $\rho\pi\pi$
in all charge combinations.  The $\omega\pi$ and $\rho\pi\pi$ mass spectra
are each a superposition of $\rho(770),~\rho(1450)$, and $\rho(1700)$
resonances.  The masses and widths of the resonances are fixed.
We adjust the parameters by comparing distributions in simulated data,
including the effects of detector acceptance and resolution,
and in real data with estimated
tau and non-tau backgrounds subtracted.  Figures~\ref{Ftwopi}
and \ref{Ffourpi} illustrate the goodness of the fit.

We compute the resolution function ${\cal R}_i$ separately for each event.
The scale
of the spreading in $x-x_i$ and $y-y_i$, including $x,y$ correlation, is
obtained by propagating the resolution error matrices from the individual
track and shower fits.  We obtain a parametrized
non-Gaussian shape from Monte Carlo.  The width and shape lead 
to distributions
that match data for the reconstructed mass of $\pi^0\to\gamma\gamma,~
K_S^0\to\pi^+\pi^-$, and $D\to K\pi$.
The projected distributions of r.m.s. resolutions in 
hadronic mass and energy peak
at 11 and 17 MeV in the region near the kinematic endpoint.
By using the resolution function appropriate for each event instead of an
averaged one, we diminish the effect of fluctuations from poorly measured
events near the kinematic boundary.

We evaluate the integral for each event by Monte Carlo, using a GEANT
\cite{GEANT}
simulation of physical processes in the CLEO detector.  That is, we first
generate about 1 million simulated signal events using the KORALB
event generator\cite{RKORB}, tagged
$\tau^-\to\pi^-\pi^+\pi^-\pi^0\nu_\tau$, according to the distribution
${\cal F}(x,y|0)$.  Then for each observed real event $i$ and for each
assumed value of $m_{\nu_\tau}$
we form the following sum over all the Monte Carlo events that contribute:
$$P_{sig}(x_i,y_i|m_{\nu_\tau})={{\sum_{j=1}^{N_{MC}}w(x_j,y_j|m_{\nu_\tau})
 {\cal R}_i(x_i-x_j,y_i-y_j)}\over{\sum_{j=1}^{N_{MC}}
 w(x_j,y_j|m_{\nu_\tau})}}.$$
We take account of the efficiency factor $\epsilon$ by omitting
the Monte Carlo events that are not detected by the simulated CLEO
detector or recognized by the event selection criteria.
The Monte Carlo integration technique enables us to include
in $P_{sig}$ the effect of initial state radiation,
$e^+e^-\to\tau^+\tau^-\gamma$.  Radiative events have a lower effective 
$E_{beam}$ causing them to be produced with lower apparent $y$.
Some of them can be seen in Fig.\ \ref{Fxy} below the lower no-radiation
kinematic limit.  The Monte Carlo also includes the appropriate number of
events from the $K_S\to\pi^+\pi^-$ and $K^\pm$-for-$\pi^\pm$ 
substitution modes
(see Table \ref{Tmodes}).
%..............................................................

{\it Background Likelihood.}
We distinguish three types of significant background:
(a) events from the two-photon
process that are not eliminated by our transverse momentum cut,
(b) $\tau^+\tau^-$ events that do not contain our signal modes,
and (c) non-$\tau$ hadronic events from $e^+e^-\to q\overline q$
 ($q=u,d,s,c$).

The two-photon events, such as $e^+e^-\to e^+e^-\gamma^*\gamma^*,
~\gamma^*\gamma^*\to\pi^+\pi^+\pi^-\pi^-\pi^0$ in which the hadronic state
has enough transverse momentum to be accepted, form a background at low
scaled hadron energy $y$ that is difficult to model reliably.  Since this
kinematic region is insensitive to $m_{\nu_\tau}$, the best strategy
is to eliminate it from the likelihood fit.  The detector
efficiency is also less accurately modeled at low $x$ and low $y$, so we
restrict the fit to $x$$>$0.7,~$y$$>$0.7.  Within this region the two-photon
background can be neglected, and the detailed choice of boundary has
no influence on the $m_{\nu_\tau}$ limit.  The cut reduces the
number of data events used in the fit to 16,577.

Tau decays of higher or lower multiplicity can masquerade as our signal 
mode if particles escape undetected
and/or secondaries in the CsI crystal
array are misinterpreted as photons from a $\pi^0$.
We evaluate these and other mis-reconstruction effects by
Monte Carlo simulation of the response of the CLEO detector to 
12 million $\tau^+\tau^-$ events generated with the known branching
fractions.  Of the accepted data events in the fit region, 7.3\% are
from tau background.  They are mainly $\tau^-\to\pi^-\pi^+\pi^-\nu_\tau$
with a spurious $\pi^0$.

Although most hadronic $e^+e^-\to q\overline q$ annihilation events
are rejected by our one-versus-three charged track
criterion, some of them can survive.  The $q\overline q$ contamination in
our data sample was evaluated by a 36-million event Monte Carlo
simulation using the LUND\cite{RLUND} generator.
The simulation has been extensively tuned to produce results that agree with
experiment.  In particular, we have verified the agreement between data
and Monte Carlo for the
events satisfying the one-versus-three topology but having a tag with
an energy that could not come from $\tau$ decay, and for the events
that have $x$ values well above the kinematic limit for the signal $\tau$
mode.  The $q\overline q$ background accounts for 3.1\% of the accepted
data events in the fit region.

The calculation of the background contribution $P_{bkg}$ to the
individual event likelihood is similar to the calculation of $P_{sig}$.
However, since the expected background event
distribution (the analog of ${\cal F}\times\epsilon$)
can be expressed only in terms of observed
$x_i,y_i$, it is not appropriate to integrate over the 
experimental resolution
function; its effect is already contained in the distribution.
As the distribution is not an analytic function,
but a collection of simulated events with a rather smooth distribution,
we approximate the value of $P_{bkg}(x_i,y_i)$ for the $i$-th
data event by the number (appropriately weighted)
of $\tau^+\tau^-$ and $q\overline q$ background Monte Carlo events
per unit area in the $x,y$ vicinity of $x_i,y_i$.
This $P_{bkg}$ is of course independent of $m_{\nu_\tau}$.

We sum the signal and background likelihoods for each event and take the
product over all events in the region
$x>0.7$, $y>0.7$ to form the net likelihood ${\cal L}$.
This is repeated for a range of assumed $m_{\nu_\tau}$ values to obtain
${\cal L}(m_{\nu_\tau})$.\footnote{A Poisson coefficient expressing the 
                                   dependence of the number of observed 
                                   events on $m_{\nu_\tau}$ 
                                   (as used in ref.~\cite{RDubos})
                                   is not used here because the large 
                                   number of events in the fit region
                                   of this work is insensitive to the 
                                   neutrino mass scale in question.}
The overall normalization of
${\cal L}(m_{\nu_\tau})$ is arbitrary.
%-------------------------------------------------------------------------
\section{Results}
\begin{figure}
\epsfig{file=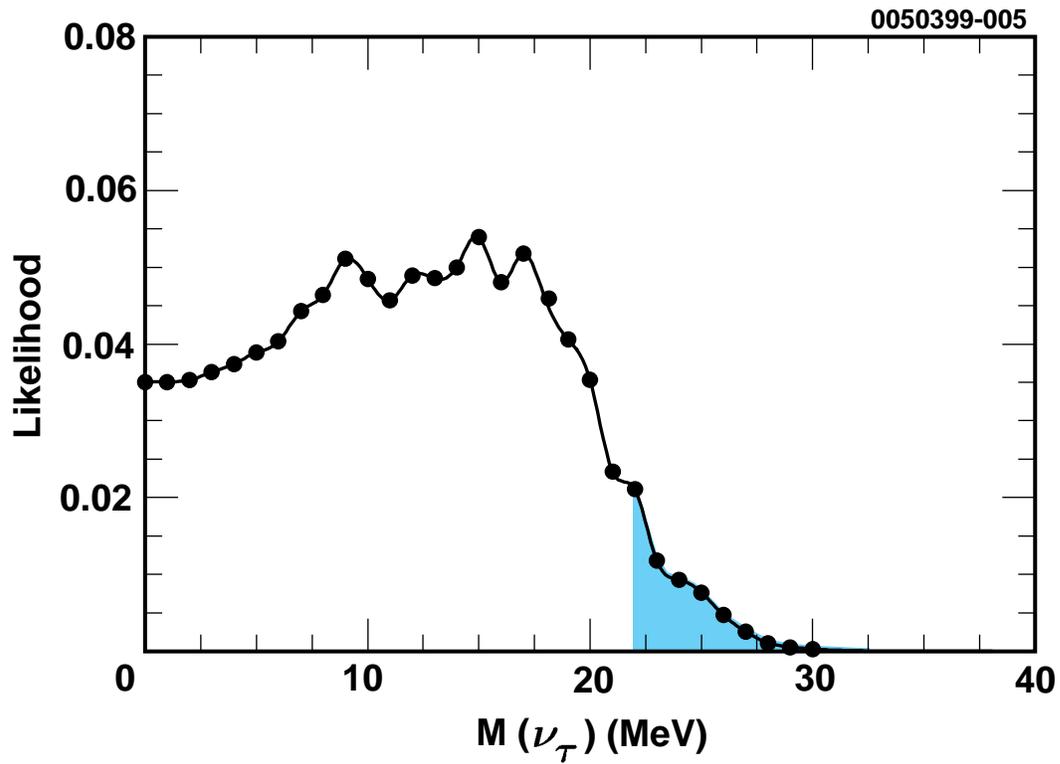,bbllx=90pt,bblly=395pt,bburx=510pt,
 bbury=690pt}
\caption{The measured likelihood evaluated for a sequence of $m_{\nu_\tau}$
values.  The curve is a cubic spline fit through the points.  
The shaded area is 5\% of the integral under the curve.}
\label{FLvsM}\end{figure}

Figure~\ref{FLvsM}(a) shows the likelihood as a function of assumed neutrino
mass.  The integral under the curve beyond 22 MeV is 5\% of the total.
Before interpreting this as a 95\% upper limit on $m_{\nu_\tau}$, however,
we have to consider systematic uncertainties that could affect the limit.

The CLEO charged particle momentum measurement scale is uncertain by about
0.05\%, and the $\pi^0$ energy scale is uncertain by 0.25\%.  This shows
up as a potential mismatch between Monte Carlo and data, resulting in a
distortion in ${\cal L}(m_{\nu_\tau})$.  
Variations in scale of these magnitudes
cause a shift of 5.0 MeV in the 95\% limit, when the two effects are
combined in quadrature.

The four-pion spectral function ${\cal F}$ was determined by varying the
contributions of the $\rho(770)$, $\rho(1450)$, and $\rho(1700)$ to match the
data below $x=0.925$, then extrapolating into the region $0.925<x<1$ 
sensitive to $m_{\nu_\tau}$.
\footnote{Note that if we were to force the
spectral function to agree with the data
all the way to the kinematic end point ($x=1$), we would get a (biased)
95\% limit of 17 MeV for $m_{\nu_\tau}$.  
This is significant in that it
represents the lowest limit one could {\sl a priori} expect to obtain with
the given statistical accuracy and background, assuming $m_{\nu_\tau}=0$.}
The resulting likelihood function (Fig.~\ref{FLvsM}) is
sensitive mainly to the parameters describing the $\rho(1700)$.
If we vary the amplitude within the experimental accuracy of our matching
and the $\rho(1700)$ mass and width within their experimental errors,
we can raise the 95\% likelihood limit by 4 MeV.

The effects of other systematic uncertainties have been similarly evaluated.
The resulting shifts of the 95\% limit on $m_{\nu_\tau}$ are
listed in Table \ref{Tsyserr}.  Assuming that the various effects are
independent, we combine the limit shifts in quadrature to get a net systematic
shift of 6.4 MeV.  Following the practice in reports of
previous experiments on $m_{\nu_\tau}$ limits, we add this shift linearly
to the raw limit from Fig.\ \ref{FLvsM}:
$$m_{\nu_\tau}<28{\rm ~MeV~~(95\%~confidence)}.$$         %28.4

An important difference between this measurement and $\nu_\tau$ mass limits
from previous $e^+e^-$ experiments
is the size of the event sample used.  We observe 543 events in the sensitive
portion of the kinematically allowed region, $x>0.925$.
This and the fact that we include background in the fit lead to a
limit that has little sensitivity to chance fluctuations in the population
of individual events near the endpoint.  Although this experiment has its own
statistical and systematic uncertainties that prevent a significant
improvement in the limit value, the analysis is quite complementary
to previous low statistics experiments and confirms their conclusions. 

Both this measurement and the previous CLEO measurement \cite{RDubos}
using $\tau\to 5\pi\nu_\tau$ show a broad likelihood maximum near the
higher end of the allowed range of the tau neutrino mass.
Since such a behavior is not unlikely even if $m_{\nu_\tau}=0$ 
(as verified by Monte Carlo experiments), we do not
regard it as significant.  It does imply, however, that combining the
results of the two CLEO measurements to make a joint likelihood curve
does not significantly improve the mass limit.
%------------------------------------------------------------------------
\section{Acknowledgments}
We gratefully acknowledge the effort of the CESR staff in providing us with
excellent luminosity and running conditions.
J.R. Patterson and I.P.J. Shipsey thank the NYI program of the NSF,
M. Selen thanks the PFF program of the NSF,
M. Selen and H. Yamamoto thank the OJI program of DOE,
J.R. Patterson, K. Honscheid, M. Selen and V. Sharma
thank the A.P. Sloan Foundation,
M. Selen and V. Sharma thank Research Corporation,
S. von Dombrowski thanks the Swiss National Science Foundation,
and H. Schwarthoff thanks the Alexander von Humboldt Stiftung for support.
This work was supported by the National Science Foundation, the
U.S. Department of Energy, and the Natural Sciences and Engineering Research
Council of Canada.

\vfill\eject

\begin{table}
\caption{95\% confidence upper limits on $m_{\nu_\tau}$ obtained from
kinematics of $\tau$ decays.}
\label{Tlimits}
\begin{center}\begin{tabular}{lccl}
Experiment & Ref.           & Decay             & MeV     \\
\hline\\
ALEPH      & ref. \cite{RDCXXIX} & $3\pi^\pm$        & 25.7      \\  %25.7
           &               & $5\pi^\pm(\pi^0)$  & 23.1      \\  %23.1
OPAL       & ref. \cite{RDCLXV}  & $5\pi^\pm$        & 43.2      \\  %43.2
           &               & $3\pi^\pm$         & 35.3      \\  %35.3
ARGUS      & ref. \cite{RDCLXVI} & $5\pi^\pm$        & 31      \\
CLEO       & ref. \cite{RDubos}  & $5\pi^\pm,3\pi^\pm2\pi^0$ & 30 \\
           & this analysis & $3\pi^\pm\pi^0$    & 28      \\  %28.4
\end{tabular}\end{center}\end{table}

\begin{table}
\caption{Fraction of the $\tau\to h^-h^+h^-\pi^0\nu_\tau$ data contributed
by each of the signal modes, 
based on known branching fractions \pcite{Rpdg}.}
\label{Tmodes}
\begin{center}\begin{tabular}{lr}
Decay mode                     & \% of signal \\
\hline\\
$\pi^-\pi^+\pi^-\pi^0\nu_\tau$                 & 92~~ \\
$K_S^0\pi^-\pi^0\nu_\tau,~~K_S^0\to\pi^+\pi^-$ &  3~~ \\
$K^-\omega\nu_\tau,~~\omega\to\pi^+\pi^-\pi^0$ &  1.6 \\
$K^-K^+\pi^-\pi^0\nu_\tau$                     &  1.5 \\
$K^-K_S^0\pi^0\nu_\tau,~~K_S^0\to\pi^+\pi^-$   &  1.1 \\
$K^-\pi^+\pi^-\pi^0\nu_\tau$                   &  0.5 \\
$\omega\pi^-\pi^0\nu_\tau,~~\omega\to\pi^+\pi^-$& 0.2 \\
\end{tabular}\end{center}\end{table}

\begin{table}
\caption{Systematic uncertainty sources and the shifts they induce in the
$m_{\nu_\tau}$ upper limit.}
\label{Tsyserr}
\begin{center}\begin{tabular}{lr}
Source                         & MeV \\
\hline\\
$\pi^0$ energy scale            & 3.7 \\
Track momentum scale           & 3.3 \\
Spectral function              & 4.0 \\
$q\overline q$ and $\tau$ background corrections & 0.8 \\
Monte Carlo statistics         & 0.5 \\
Resolution function            & 0.4 \\
\hline
Quadrature sum                 & 6.4 \\
\end{tabular}\end{center}\end{table}


\end{document}